\documentclass[aip,apl,amsmath,amssymb,reprint,groupedaddress]{revtex4-1}

\usepackage{graphicx}
\usepackage{dcolumn}
\usepackage{bm}

\newcommand{\ket}[1]{\ensuremath{\left|#1\right>}}

\newcommand{\yb}{$^{171}$Yb$^+$ }
\newcommand{\sfz}{$^2$S$_{1/2}{\left|F=0, m_f=0\right>}$ }
\newcommand{\sfo}{$^2$S$_{1/2}{\left|F=1\right>}$ }

\newcommand{\sfoc}{$^2$S$_{1/2}{\left|F=1,m_f=0\right>}$ }

\newcommand{\pfz}{$^2$P$_{1/2}{\left|F=0\right>}$ }
\newcommand{\pfo}{$^2$P$_{1/2}{\left|F=1\right>}$ }

\begin{document}
\title{Individual addressing of trapped \yb ion qubits using a MEMS-based beam steering system}
\author{S. Crain}
\author{E. Mount}
\author{S. Baek}
\author{J. Kim}
	\email{jungsang@duke.edu}
\affiliation{Fitzpatrick Institute for Photonics, Electrical and Computer Engineering Department, Duke University, Durham, North Carolina 27708, USA}
\date{\today}

\begin{abstract}
The ability to individually manipulate the increasing number of qubits is one of the many challenges towards scalable quantum information processing with trapped ions. Using micro-mirrors fabricated with micro-electromechanical systems (MEMS) technology, we focus laser beams on individual ions in a linear chain and steer the focal point in two dimensions. We demonstrate sequential single qubit gates on multiple \yb qubits and characterize the gate performance using quantum state tomography. Our system features negligible crosstalk to neighboring ions ($< 3\times 10^{-4}$), and switching speed comparable to typical single qubit gate times ($<$ 2 $\mu$s). 
\end{abstract}

\maketitle
Trapped ions are one of the leading candidates to be used as quantum bits (qubits) for quantum information processing (QIP) due to their high fidelity logic operations available\cite{Wineland1998a,Blatt2008a}.  Single and multi-qubit gate operations have been performed on trapped ions using lasers or microwave fields\cite{Mount2013, Leibfried2003, Ospelkaus2011, Allcock2013}.  Extending these operations over multiple qubits with individual qubit addressing has been accomplished by using acousto/electro-optic modulators to steer laser beams\cite{Kaler2003,Yavuz2006}, or engineering magnetic field gradients at the ion location\cite{Johanning2009}.  As the number of qubits in a QIP system increases, one needs a scalable approach to manipulate each individual qubit\cite{Seidelin2006, Kim2005, Nagerl1998}.  In this letter we utilize a two-dimensional beam steering system based on MEMS mirrors\cite{Knoernschild2010} to steer laser beams across a chain of trapped \yb ions.  We characterize the crosstalk of the laser beam onto neighboring sites, measure the switching speed between two target sites, and characterize the fidelity of single qubit gate operations performed on a pair of trapped ion qubits.        

\begin{figure}[b]
	\includegraphics[width=\columnwidth]{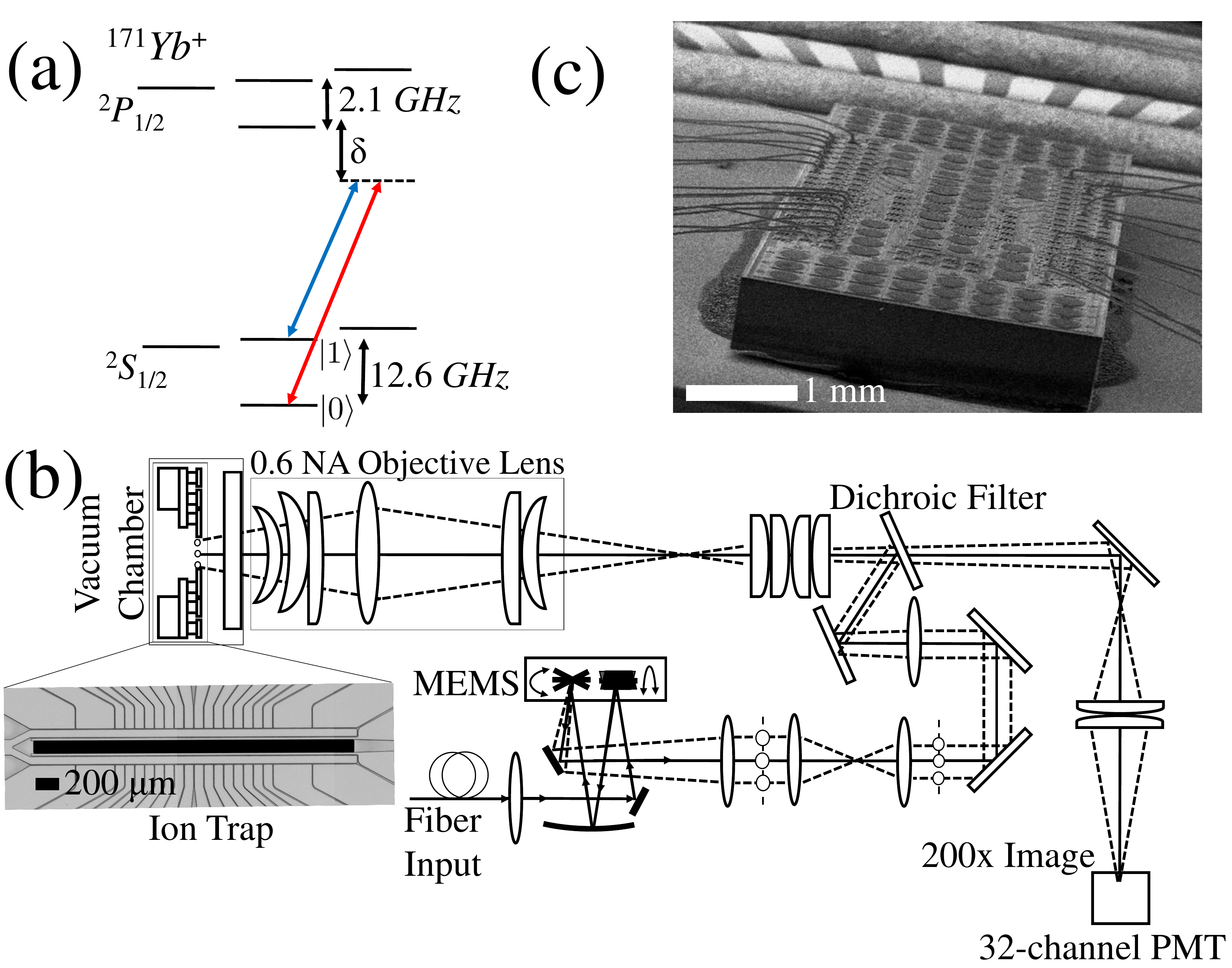}
	\caption{\label{fig:experimentalSetup} (Color online) (a) Energy level diagram of \yb depicting the two photon Raman transition for single qubit operations (not to scale).  (b) Schematic of optical setup (both imaging and MEMS beam paths) as well as a microscope image of the ion trap highlighting the slot (the dark region) and control electrodes.  (c) SEM image of a MEMS chip.}
\end{figure}

Our qubit is defined by the two hyperfine ground states of the \yb ion, \ket{0} $\equiv$ \sfz and \ket{1} $\equiv$ \sfoc (Figure \ref{fig:experimentalSetup} (a)).  Doppler cooling, optical pumping for qubit initialization, resonant scattering for qubit detection, and repump from the $^2$D$_{3/2}$ level are all performed using continuous wave (CW) external cavity diode lasers\cite{Olmschenk2007}.  The ion is prepared in the \ket{0} state by applying light resonant with the \sfo $\rightarrow$ \pfo transition for 20 $\mu$s\cite{Olmschenk2007}.  For state detection, the ion is exposed to light resonant with the \sfo $\rightarrow$ \pfz\cite{Olmschenk2007, Noek2013} transition.  If the ion is in the \ket{0} state, the light is not resonant with any allowed transitions and the ion does not scatter any photons.  If the ion is in the \ket{1} state, the ion will scatter photons at a known rate due to the cycling transition between the \sfo and \pfz states.  

Single quit rotations between the \ket{0} and \ket{1} state can be driven by a stimulated Raman transition using two laser beams with a frequency difference resonant with the hyperfine splitting, $\Delta_{hf}$ = 12.6 GHz (Figure \ref{fig:experimentalSetup}(a)) \cite{Mount2013}.  For our experiments, frequency combs from phase-locked ultra-fast lasers are used to drive Raman transitions\cite{Hayes2010}.  We use a picosecond titanium-sapphire laser with a center frequency close to 752 nm and a 76 MHz repetition rate.  The laser frequency is doubled to produce a center wavelength of 376 nm, which is red-detuned from the ion's \sfo $\rightarrow$ \pfz resonance by approximately $\delta$ = 14 THz.  An acousto-optic modulator (AOM) is used to shift the two frequency combs relative to each other.  We use a direct digital synthesizer (DDS) to set the modulation frequencies, allowing for digital control of the frequency, amplitude, and phase of each comb.  The frequency shift is stabilized so that the frequency difference between the two Raman beams plus an integer multiple of the repetition rate spans the hyperfine splitting\cite{Mount2013, Hayes2010}.  Co-propagating Raman beams are implemented by driving an AOM with two modulation frequencies and are circularly polarized with respect to the quantization axis defined by the magnetic field.      

The ion trap is a micro fabricated radio frequency Paul trap designed and fabricated by Sandia National Laboratories (inset in Figure \ref{fig:experimentalSetup}(b))\cite{Mount2013}, mounted in a 4.5-inch ultra-high vacuum spherical octagon chamber (Kimball Physics).  Chains of ions are trapped 80 $\mu$m above the trap surface.  A custom re-entrant viewport allows us to position an anti-reflection coated window 11 mm from the trap surface, and we image the ion through the re-entrant viewport with a custom 0.6 NA objective lens (Photon Gear, Inc.)\cite{Noek2013}.  Further imaging optics magnify the ion image by $\sim$200x (see Figure \ref{fig:experimentalSetup}(a)) onto a 32-channel segmented photomultiplier tube (PMT) with custom readout circuitry.  Each segment of the PMT detects light scattered from a single ion in the chain, enabling parallel readout of multiple qubits in the chain. 

Beam steering of the Raman laser is accomplished by using two one-dimensional MEMS mirrors\cite{Kim2007} tilting in orthogonal directions.  The mirrors are fabricated on a single substrate (Figure \ref{fig:experimentalSetup}(c)) using Sandia's SUMMiT V process\cite{Smith1998} allowing for a single system to be scaled to multiple beams across a variety of wavelengths\cite{Knoernschild2008,Knoernschild2009}.  The mirrors consist of a 125 $\mu$m radius polysilicon plate with a 30 nm thick aluminum coating which allows for maximum reflectance at 376 nm wavelength while minimizing the stress that induces curvature on the plate.  The beam waist of the incoming Raman beam is focused onto the first of the MEMS mirrors, and a folded 2$f$-2$f$ imaging system projects the beam waist onto the second MEMS mirror\cite{Knoernschild2008}.  By applying an actuation voltage between the grounded mirror plate and underlying electrode, the mirrors can be tilted by an angle $\theta$ (beam tilt angle of $2\theta$).  The voltage is supplied by a custom digital-to-analog converter (DAC) circuit followed by a fast high-voltage amplifier.  A lens placed a focal length $f$ away from the second mirror translates the tilts of the laser beam reflecting off the mirrors to a vertical and horizontal shift of $2\theta*f$ in the Fourier plane.  The resulting output beam is then demagnified and projected onto the ion location through the same 0.6 NA lens used as the imaging system.  This is accomplished by using a dichroic filter that reflects the Raman beam (at 376 nm) and transmits the light scattered by the ion (at 369.5 nm).  The Raman beam is perpendicular to the trap surface, allowing it to pass through the slot in the middle of the ion trap\cite{Mount2013}.  

\begin{figure} 
	\includegraphics[width=\columnwidth]{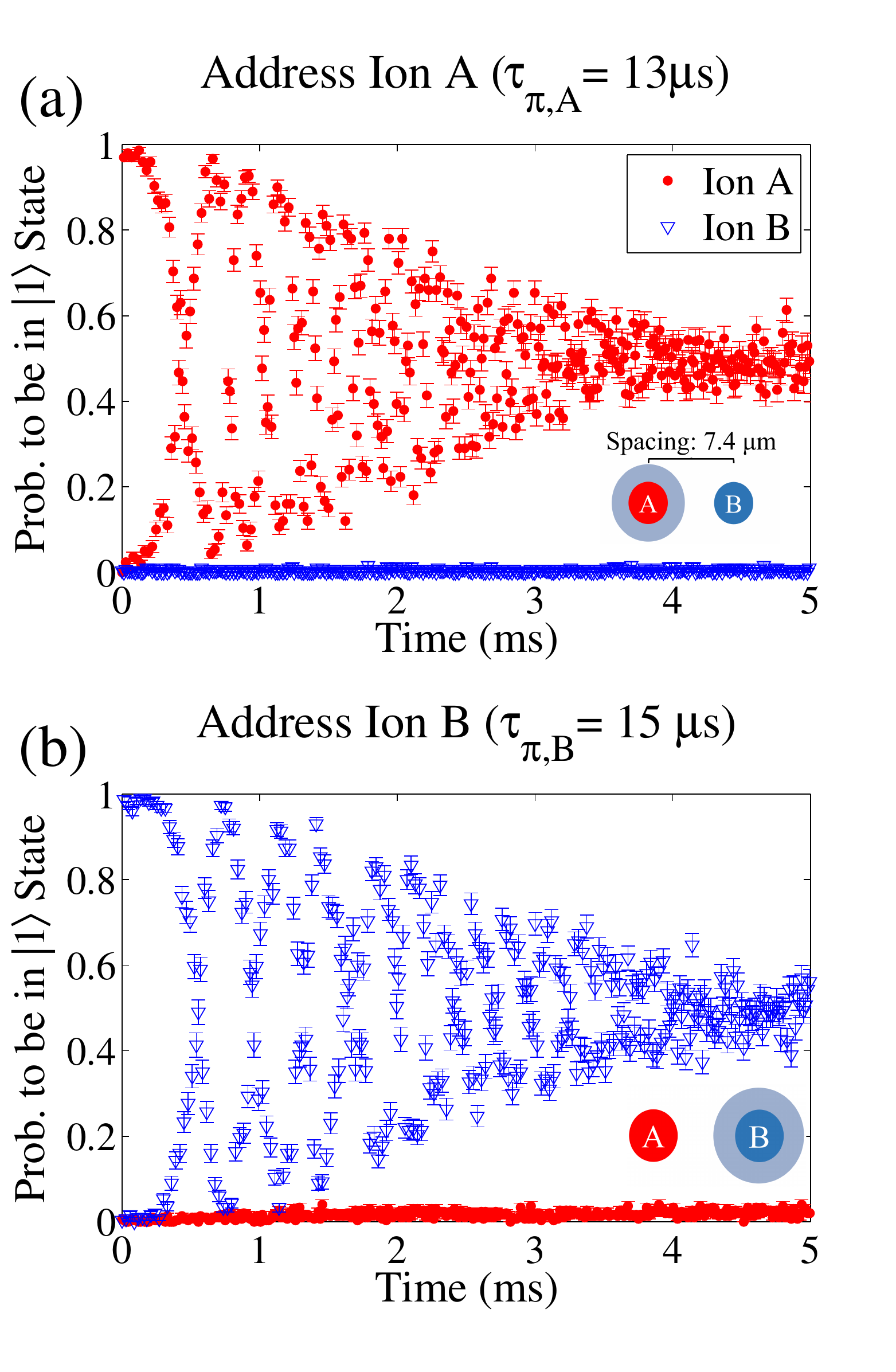}
	\caption{\label{fig:crosstalk}(Color online) (a) The Raman beam is directed at ion A while ion B is prepared in the \ket{0}.  The step size for the Raman beam duration is initially set to $\tau_{\pi,A} =$ 13 $\mu$s, but because of intensity drift at the ion, $\tau_{\pi,A}$ varies slightly throughout the experiment.  Therefore, the plot for ion A shows the envelope of the Rabi oscillations.  (b) The Raman beam is directed at ion B, and ion A is prepared in the \ket{0} state.}
\end{figure}

For the first experiment, we characterize the amount of crosstalk from the Raman beam on neighboring ion sites separated by approximately 7.4 $\mu$m.  The two ions are first initialized to the \ket{0} state.  The Raman beam is then directed onto one of the two ions (site A) for a duration $T$.  The resulting state of the two ions is then determined by state-dependent fluorescence.  The same experiment is then repeated with the Raman beam directed onto the second of the two ions (site B).  Figures \ref{fig:crosstalk}(a) and \ref{fig:crosstalk}(b) show the probability of the ions to be in the \ket{1} state as a function of the duration of the Raman beam, $T$.  The amount of time needed in order to transition the ion from the \ket{0} state to the \ket{1} state is $\tau_{\pi,A}$ = 13 $\mu$s and $\tau_{\pi,B}$ = 15 $\mu$s for ion A and ion B, respectively.  The variation between $\tau_{\pi,A}$ and $\tau_{\pi,B}$ is due to intensity variation of the Raman beams between the two experiments.  After 5 ms ($\sim$350 $\tau_{\pi,A(B)}$), the probability of the target ion to be in the \ket{1} state averages to 0.5 due to the laser intensity drift, while that for the neighboring ion stays low ($\sim$0.005 for ion B $\sim$0.024 for ion A).  From the ratio of the Rabi frequencies between the target and neighboring ion, we estimate that the amount of intensity crosstalk at the neighboring ion is $1.3\times 10^{-4}$ on ion B and $2.9\times 10^{-4}$ on ion A.

In order to estimate the beam waist at the ion location, the Raman beam is parked at halfway between the location of the two ions, and $\tau_{\pi,A}$ and $\tau_{\pi,B}$ are measured.  Leaving the Raman beam at the same location, a single ion is loaded to the middle position (location C) and its corresponding $\tau_{\pi,C}$ is measured.  The power of the Raman beam and the spacing of the two ions is known, so the beam waist can be calculated from the ratio of $\tau_{\pi,A}$ ($\tau_{\pi,B}$)  and $\tau_{\pi,C}$ assuming a Gaussian beam.  From this measurement, we estimate the beam waist to be $\sim$3.3 $\mu$m, which should lead to an intensity crosstalk of $<5\times 10^{-5}$.  An imperfect Gaussian beam shape and unwanted scatter ($< -30$ dB) could be the cause of the difference between the measured and estimated crosstalk. 

\begin{figure}
\includegraphics[width=\columnwidth, clip=true, trim=0 0 0 40 ]{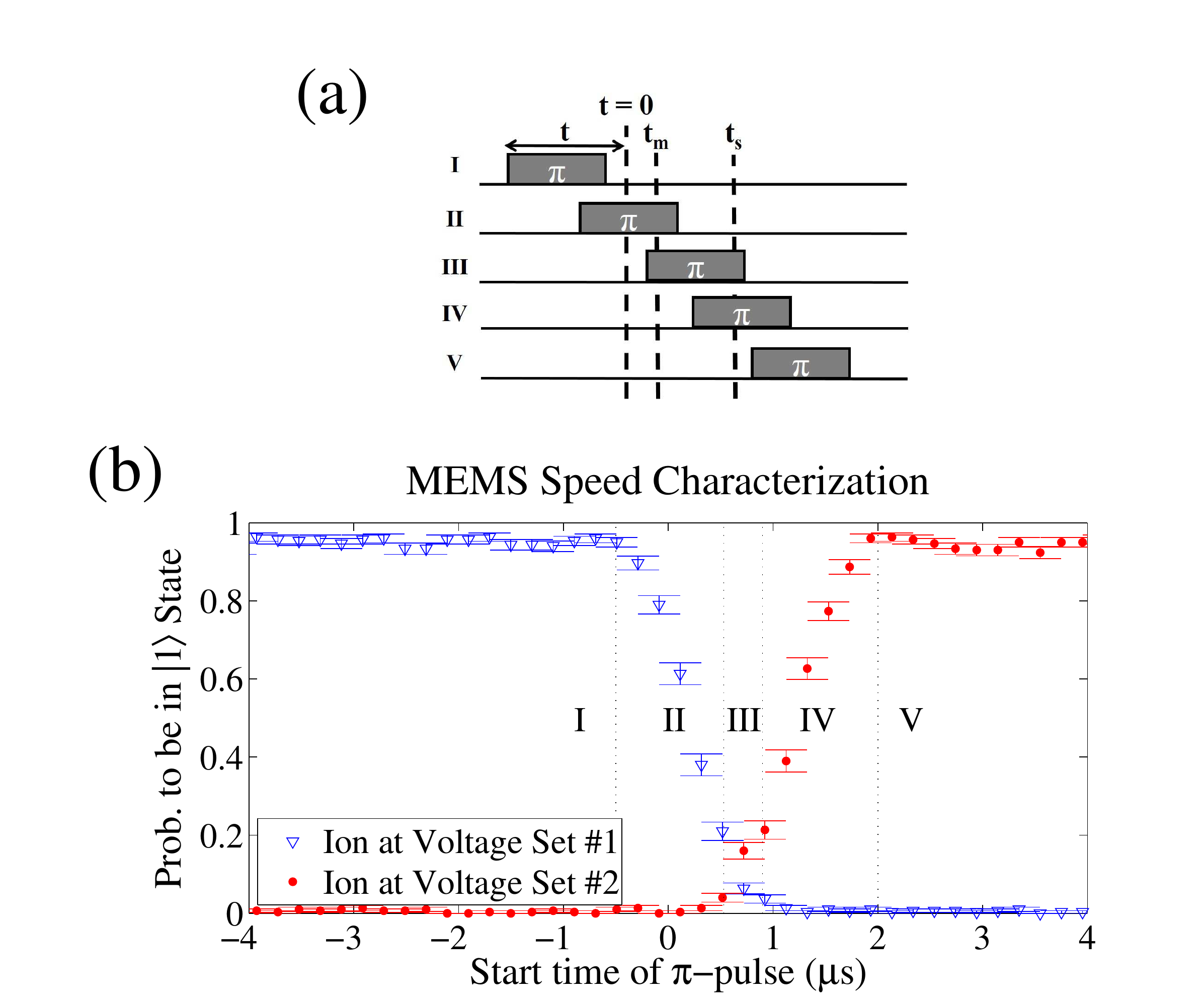}
\caption{\label{fig:memsSwitchSpeed}(Color online) (a) Timing diagram of the MEMS speed characterization experiment.  The applied voltages for the MEMS electrodes are switched at time $t = 0$.  The response time of the MEMS mirrors is characterized by varying the start time of the $\pi$-pulse, $t$. Time $t_m$ ($\sim$0.9 $\mu$s) denotes the delay of the MEMS mirrors' mechanical response to the applied voltage and $t_s$ ($\sim$2 $\mu$s) denotes time needed for a full $\pi$-pulse incident on the ion after switching the MEMS.  (b) Response time of the MEMS mirrors when switched between two neighboring ion sites.  Region I is the case where the $\pi$-pulse finishes before $t_m$, which results in a full $\pi$-pulse on the ion site corresponding to voltage set $\#$1.  Region II corresponds to when the $\pi$-pulse extends past $t_m$ but ends before $t_s$, resulting in a partial $\pi$-pulse experienced by the ion site corresponding to voltage set $\#$1.  Region III corresponds to when the $\pi$-pulse begins before $t_m$ and ends after $t_s$, which results in a partial $\pi$-pulse on both ion sites corresponding to both voltage sets.  Region IV is the case when the $\pi$-pulse is started after $t_m$ but before $t_s$, which results in the ion site corresponding to voltage set $\#$2 experiencing a partial $\pi$-pulse.  Region V is the case where the $\pi$-pulse starts after $t_s$, resulting in a full $\pi$-pulse on the ion site corresponding to voltage set $\#$2.  The time for the mirrors to switch completely from one ion site to the other is approximately 1.1 $\mu$s.}
\end{figure}

In order to characterize the switching time between two neighboring ion sites, we measure the response time of the MEMS by switching the Raman beam on and off of a single ion.  First, we align the Raman beam on the single ion using voltage set $\#$1. We also identify voltage set $\#$2 that shifts the Raman beam to a neighboring site, separated by the distance between two trapped ions ($\sim$7.4 $\mu$m).  After initializing the ion to the \ket{0} state, the pulsed laser is turned on for $\tau_{\pi,1}$ = $1.5$ $\mu$s at time $t$.  The voltages applied to the MEMS electrodes are switched from set $\#$1 to set $\#$2 at $t = 0$ (Figure\ref{fig:memsSwitchSpeed}(a)).  The state of the ion is then measured by state-dependent fluorescence after the Raman $\pi$-pulse is complete.  The response time of the MEMS beam to move off the ion can be inferred by monitoring the rate at which the Raman transition is suppressed (blue triangles in Figure \ref{fig:memsSwitchSpeed}(b)).  Next, the experiment is repeated by re-aligning the Raman beam to hit the ion when voltage set $\#$2 is applied to the MEMS electrodes ($\tau_{\pi,2}$ = $1.3$ $\mu$s). For this case, the beam starts off the ion when voltage set $\#$1 is used, and then comes onto the ion as the voltages are triggered to set $\#$2. The rate at which the Raman transition is activated (red dots in Figure \ref{fig:memsSwitchSpeed}(b)) indicates the time scale over which the MEMS beam moves onto the ion.  The total time it takes to switch the MEMS mirrors to a neighboring ion site can be calculated from the time between the cases where both ion sites experience a full $\pi$-pulse, less $\tau_{\pi,1}$ (or $\tau_{\pi,2}$).  The calculated switching time from this data is approximately 1.1 $\mu$s.  The slope of the response curve is limited by the $\tau_{\pi,1}$ we can achieve with our current optical power and beam waist.  

\begin{table}
\setlength{\tabcolsep}{15pt}
\renewcommand{\arraystretch}{1.25}
\begin{tabular}{ | c | c | } 
	\hline 
	Gate A, Gate B & Gate Fidelities\\ \hline
	$R_x(\frac{\pi}{2}),\mathcal{I}$ & 0.991(4), 0.9974(5) \\ 
	$\mathcal{I},R_x(\frac{\pi}{2})$ & 0.9972(5), 0.991(4) \\ 
	$R_x(\pi),R_y(\frac{\pi}{2})$ & 0.989(1), 0.992(4) \\
	$R_x(\frac{\pi}{2}),R_x(\pi)$ & 0.989(4), 0.9913(9) \\ 
	$R_x(\frac{\pi}{2}),R_x(\frac{\pi}{2})$ & 0.991(4), 0.992(4) \\ 
	$R_x(\frac{\pi}{2}),R_y(\frac{\pi}{2})$ & 0.990(4), 0.992(4) \\ 
	$R_y(\frac{\pi}{2}),R_x(\frac{\pi}{2})$ & 0.990(4), 0.992(4) \\ 
	 \hline
\end{tabular}
\caption{\label{tab:tomoTable}Table of various combinations of sequential single qubit gates on a pair of trapped ions and their corresponding gate fidelities measured using quantum state tomography.} 
\end{table}

Our final experiment measures the fidelity of two sequential single qubit gates on a pair of ions using quantum state tomography.  Quantum state tomography reconstructs the full density matrix of the qubit state by projecting each qubit into the \ket{0}, $\frac{1}{\sqrt{2}}(\ket{0}+\ket{1})$, and $\frac{1}{\sqrt{2}}(\ket{0}+\imath\ket{1})$ bases\cite{Kwiat2006}.  We first prepare both qubits in the  \ket{0} state.  The MEMS mirrors then tilt to aim the addressing beam at ion A and the Raman beam is turned on for the first single qubit gate operation.  After the Raman beam is turned off, the MEMS mirrors aim the addressing beam at ion B.  Then the Raman beam is turned on for the second single qubit operation.  The Raman beam is redirected back to ion A and the qubit is rotated to one of the three measurement bases by applying the appropriate gate:  $\mathcal{I}$, $R_x(\frac{\pi}{2})$, or $R_y(\frac{\pi}{2})$ .  After rotating ion B to the same measurement basis, both qubits are measured using state-dependent fluorescence.  This sequence is performed again with the same single qubit operations for each of the other two measurement bases.  Table \ref{tab:tomoTable} shows different combinations of sequential single qubit gates on ion A and ion B along with the respective gate fidelities measured using state tomography.  The fidelity of the preparation and measurement in the \ket{0} state is 0.998 and the fidelity of measuring the \ket{1} state is 0.991 in this experiment.  The overall gate fidelities are currently dominated by our state preparation and measurement (SPAM) errors.  In order to isolate the gate fidelities from SPAM errors, one can use randomized benchmarking technique\cite{Knill2008}.   

This letter has demonstrated the individual addressability of trapped \yb ion qubits in a chain with a two-dimensional MEMS beam steering system.  The system produces low crosstalk on neighboring ion sites at a level $< 3\times 10^{-4}$ with a beam waist at the ion location of  $\sim$3.3 $\mu$m.  Fast switching speeds ($\sim$1.1 $\mu$s) comparable to the single qubit gate times allow for minimal wait times between qubit operations at different ion locations.  Sequential single qubit operations on a pair of ions were performed and were characterized with state tomography.  This demonstrates the feasibility of utilizing a MEMS-based beam steering system as a scalable and practical solution to individual addressing of trapped ions.   

 This research was funded by the Office of the Director of National Intelligence (ODNI) and Intelligence Advanced Research Projects Activity (IARPA) through the Army Research Office.        

\bibliography{references}

\begin{thebibliography}{22}%
\makeatletter
\providecommand \@ifxundefined [1]{%
 \@ifx{#1\undefined}
}%
\providecommand \@ifnum [1]{%
 \ifnum #1\expandafter \@firstoftwo
 \else \expandafter \@secondoftwo
 \fi
}%
\providecommand \@ifx [1]{%
 \ifx #1\expandafter \@firstoftwo
 \else \expandafter \@secondoftwo
 \fi
}%
\providecommand \natexlab [1]{#1}%
\providecommand \enquote  [1]{``#1''}%
\providecommand \bibnamefont  [1]{#1}%
\providecommand \bibfnamefont [1]{#1}%
\providecommand \citenamefont [1]{#1}%
\providecommand \href@noop [0]{\@secondoftwo}%
\providecommand \href [0]{\begingroup \@sanitize@url \@href}%
\providecommand \@href[1]{\@@startlink{#1}\@@href}%
\providecommand \@@href[1]{\endgroup#1\@@endlink}%
\providecommand \@sanitize@url [0]{\catcode `\\12\catcode `\$12\catcode
  `\&12\catcode `\#12\catcode `\^12\catcode `\_12\catcode `\%12\relax}%
\providecommand \@@startlink[1]{}%
\providecommand \@@endlink[0]{}%
\providecommand \url  [0]{\begingroup\@sanitize@url \@url }%
\providecommand \@url [1]{\endgroup\@href {#1}{\urlprefix }}%
\providecommand \urlprefix  [0]{URL }%
\providecommand \Eprint [0]{\href }%
\providecommand \doibase [0]{http://dx.doi.org/}%
\providecommand \selectlanguage [0]{\@gobble}%
\providecommand \bibinfo  [0]{\@secondoftwo}%
\providecommand \bibfield  [0]{\@secondoftwo}%
\providecommand \translation [1]{[#1]}%
\providecommand \BibitemOpen [0]{}%
\providecommand \bibitemStop [0]{}%
\providecommand \bibitemNoStop [0]{.\EOS\space}%
\providecommand \EOS [0]{\spacefactor3000\relax}%
\providecommand \BibitemShut  [1]{\csname bibitem#1\endcsname}%
\let\auto@bib@innerbib\@empty
\bibitem [{\citenamefont {Wineland}\ \emph {et~al.}(1998)\citenamefont
  {Wineland}, \citenamefont {Monroe}, \citenamefont {Itano}, \citenamefont
  {King}, \citenamefont {Leibfried}, \citenamefont {Meekhof}, \citenamefont
  {Myatt},\ and\ \citenamefont {Wood}}]{Wineland1998a}%
  \BibitemOpen
  \bibfield  {author} {\bibinfo {author} {\bibfnamefont {D.~J.}\ \bibnamefont
  {Wineland}}, \bibinfo {author} {\bibfnamefont {C.}~\bibnamefont {Monroe}},
  \bibinfo {author} {\bibfnamefont {W.~M.}\ \bibnamefont {Itano}}, \bibinfo
  {author} {\bibfnamefont {B.~E.}\ \bibnamefont {King}}, \bibinfo {author}
  {\bibfnamefont {D.}~\bibnamefont {Leibfried}}, \bibinfo {author}
  {\bibfnamefont {D.~M.}\ \bibnamefont {Meekhof}}, \bibinfo {author}
  {\bibfnamefont {C.}~\bibnamefont {Myatt}}, \ and\ \bibinfo {author}
  {\bibfnamefont {C.}~\bibnamefont {Wood}},\ }\href@noop {} {\bibfield
  {journal} {\bibinfo  {journal} {Fortschritte Der Physik-Progress of Physics}\
  }\textbf {\bibinfo {volume} {46}},\ \bibinfo {pages} {363} (\bibinfo {year}
  {1998})}\BibitemShut {NoStop}%
\bibitem [{\citenamefont {Blatt}\ and\ \citenamefont
  {Wineland}(2008)}]{Blatt2008a}%
  \BibitemOpen
  \bibfield  {author} {\bibinfo {author} {\bibfnamefont {R.}~\bibnamefont
  {Blatt}}\ and\ \bibinfo {author} {\bibfnamefont {D.}~\bibnamefont
  {Wineland}},\ }\href@noop {} {\bibfield  {journal} {\bibinfo  {journal}
  {Nature}\ }\textbf {\bibinfo {volume} {453}},\ \bibinfo {pages} {1008}
  (\bibinfo {year} {2008})}\BibitemShut {NoStop}%
\bibitem [{\citenamefont {Mount}\ \emph {et~al.}(2013)\citenamefont {Mount},
  \citenamefont {Baek}, \citenamefont {Maunz}, \citenamefont {Blain},
  \citenamefont {Stick}, \citenamefont {Gaultney}, \citenamefont {Crain},
  \citenamefont {Noek}, \citenamefont {Kim},\ and\ \citenamefont
  {Kim}}]{Mount2013}%
  \BibitemOpen
  \bibfield  {author} {\bibinfo {author} {\bibfnamefont {E.}~\bibnamefont
  {Mount}}, \bibinfo {author} {\bibfnamefont {S.}~\bibnamefont {Baek}},
  \bibinfo {author} {\bibfnamefont {P.}~\bibnamefont {Maunz}}, \bibinfo
  {author} {\bibfnamefont {M.}~\bibnamefont {Blain}}, \bibinfo {author}
  {\bibfnamefont {D.}~\bibnamefont {Stick}}, \bibinfo {author} {\bibfnamefont
  {D.}~\bibnamefont {Gaultney}}, \bibinfo {author} {\bibfnamefont
  {S.}~\bibnamefont {Crain}}, \bibinfo {author} {\bibfnamefont
  {R.}~\bibnamefont {Noek}}, \bibinfo {author} {\bibfnamefont {T.}~\bibnamefont
  {Kim}}, \ and\ \bibinfo {author} {\bibfnamefont {J.}~\bibnamefont {Kim}},\
  }\href@noop {} {\bibfield  {journal} {\bibinfo  {journal} {New Journal of
  Physics}\ }\textbf {\bibinfo {volume} {15}} (\bibinfo {year}
  {2013})}\BibitemShut {NoStop}%
\bibitem [{\citenamefont {Leibfried}\ \emph {et~al.}(2003)\citenamefont
  {Leibfried}, \citenamefont {DeMarco}, \citenamefont {Meyer}, \citenamefont
  {Lucas}, \citenamefont {Barrett}, \citenamefont {Britton}, \citenamefont
  {Itano}, \citenamefont {Jelenkovic}, \citenamefont {Langer}, \citenamefont
  {Rosenband},\ and\ \citenamefont {Wineland}}]{Leibfried2003}%
  \BibitemOpen
  \bibfield  {author} {\bibinfo {author} {\bibfnamefont {D.}~\bibnamefont
  {Leibfried}}, \bibinfo {author} {\bibfnamefont {B.}~\bibnamefont {DeMarco}},
  \bibinfo {author} {\bibfnamefont {V.}~\bibnamefont {Meyer}}, \bibinfo
  {author} {\bibfnamefont {D.}~\bibnamefont {Lucas}}, \bibinfo {author}
  {\bibfnamefont {M.}~\bibnamefont {Barrett}}, \bibinfo {author} {\bibfnamefont
  {J.}~\bibnamefont {Britton}}, \bibinfo {author} {\bibfnamefont {W.~M.}\
  \bibnamefont {Itano}}, \bibinfo {author} {\bibfnamefont {B.}~\bibnamefont
  {Jelenkovic}}, \bibinfo {author} {\bibfnamefont {C.}~\bibnamefont {Langer}},
  \bibinfo {author} {\bibfnamefont {T.}~\bibnamefont {Rosenband}}, \ and\
  \bibinfo {author} {\bibfnamefont {D.~J.}\ \bibnamefont {Wineland}},\
  }\href@noop {} {\bibfield  {journal} {\bibinfo  {journal} {Nature}\ }\textbf
  {\bibinfo {volume} {422}},\ \bibinfo {pages} {412} (\bibinfo {year}
  {2003})}\BibitemShut {NoStop}%
\bibitem [{\citenamefont {Ospelkaus}\ \emph {et~al.}(2011)\citenamefont
  {Ospelkaus}, \citenamefont {Warring}, \citenamefont {Colombe}, \citenamefont
  {Brown}, \citenamefont {Amini}, \citenamefont {Leibfried},\ and\
  \citenamefont {Wineland}}]{Ospelkaus2011}%
  \BibitemOpen
  \bibfield  {author} {\bibinfo {author} {\bibfnamefont {C.}~\bibnamefont
  {Ospelkaus}}, \bibinfo {author} {\bibfnamefont {U.}~\bibnamefont {Warring}},
  \bibinfo {author} {\bibfnamefont {Y.}~\bibnamefont {Colombe}}, \bibinfo
  {author} {\bibfnamefont {K.~R.}\ \bibnamefont {Brown}}, \bibinfo {author}
  {\bibfnamefont {J.~M.}\ \bibnamefont {Amini}}, \bibinfo {author}
  {\bibfnamefont {D.}~\bibnamefont {Leibfried}}, \ and\ \bibinfo {author}
  {\bibfnamefont {D.~J.}\ \bibnamefont {Wineland}},\ }\href@noop {} {\bibfield
  {journal} {\bibinfo  {journal} {Nature}\ }\textbf {\bibinfo {volume} {476}},\
  \bibinfo {pages} {181} (\bibinfo {year} {2011})}\BibitemShut {NoStop}%
\bibitem [{\citenamefont {Allcock}\ \emph {et~al.}(2013)\citenamefont
  {Allcock}, \citenamefont {Harty}, \citenamefont {Ballance}, \citenamefont
  {Keitch}, \citenamefont {Linke}, \citenamefont {Stacey},\ and\ \citenamefont
  {Lucas}}]{Allcock2013}%
  \BibitemOpen
  \bibfield  {author} {\bibinfo {author} {\bibfnamefont {D.~T.~C.}\
  \bibnamefont {Allcock}}, \bibinfo {author} {\bibfnamefont {T.~P.}\
  \bibnamefont {Harty}}, \bibinfo {author} {\bibfnamefont {C.~J.}\ \bibnamefont
  {Ballance}}, \bibinfo {author} {\bibfnamefont {B.~C.}\ \bibnamefont
  {Keitch}}, \bibinfo {author} {\bibfnamefont {N.~M.}\ \bibnamefont {Linke}},
  \bibinfo {author} {\bibfnamefont {D.~N.}\ \bibnamefont {Stacey}}, \ and\
  \bibinfo {author} {\bibfnamefont {D.~M.}\ \bibnamefont {Lucas}},\ }\href@noop
  {} {\bibfield  {journal} {\bibinfo  {journal} {Applied Physics Letters}\
  }\textbf {\bibinfo {volume} {102}} (\bibinfo {year} {2013})}\BibitemShut
  {NoStop}%
\bibitem [{\citenamefont {Schmidt-Kaler}\ \emph {et~al.}(2003)\citenamefont
  {Schmidt-Kaler}, \citenamefont {Kaffner}, \citenamefont {Riebe},
  \citenamefont {Gulde}, \citenamefont {Lancaster}, \citenamefont {Deuschle},
  \citenamefont {Becher}, \citenamefont {Roos}, \citenamefont {Eschner},\ and\
  \citenamefont {Blatt}}]{Kaler2003}%
  \BibitemOpen
  \bibfield  {author} {\bibinfo {author} {\bibfnamefont {F.}~\bibnamefont
  {Schmidt-Kaler}}, \bibinfo {author} {\bibfnamefont {H.}~\bibnamefont
  {Kaffner}}, \bibinfo {author} {\bibfnamefont {M.}~\bibnamefont {Riebe}},
  \bibinfo {author} {\bibfnamefont {S.}~\bibnamefont {Gulde}}, \bibinfo
  {author} {\bibfnamefont {G.}~\bibnamefont {Lancaster}}, \bibinfo {author}
  {\bibfnamefont {C.}~\bibnamefont {Deuschle}}, \bibinfo {author}
  {\bibfnamefont {C.}~\bibnamefont {Becher}}, \bibinfo {author} {\bibfnamefont
  {C.}~\bibnamefont {Roos}}, \bibinfo {author} {\bibfnamefont {J.}~\bibnamefont
  {Eschner}}, \ and\ \bibinfo {author} {\bibfnamefont {R.}~\bibnamefont
  {Blatt}},\ }\href@noop {} {\bibfield  {journal} {\bibinfo  {journal}
  {Nature}\ }\textbf {\bibinfo {volume} {422}},\ \bibinfo {pages} {408}
  (\bibinfo {year} {2003})}\BibitemShut {NoStop}%
\bibitem [{\citenamefont {Yavuz}\ \emph {et~al.}(2006)\citenamefont {Yavuz},
  \citenamefont {Kulatunga}, \citenamefont {Urban}, \citenamefont {Johnson},
  \citenamefont {Proite}, \citenamefont {Henage}, \citenamefont {Walker},\ and\
  \citenamefont {Saffman}}]{Yavuz2006}%
  \BibitemOpen
  \bibfield  {author} {\bibinfo {author} {\bibfnamefont {D.}~\bibnamefont
  {Yavuz}}, \bibinfo {author} {\bibfnamefont {P.}~\bibnamefont {Kulatunga}},
  \bibinfo {author} {\bibfnamefont {E.}~\bibnamefont {Urban}}, \bibinfo
  {author} {\bibfnamefont {T.}~\bibnamefont {Johnson}}, \bibinfo {author}
  {\bibfnamefont {N.}~\bibnamefont {Proite}}, \bibinfo {author} {\bibfnamefont
  {T.}~\bibnamefont {Henage}}, \bibinfo {author} {\bibfnamefont
  {T.}~\bibnamefont {Walker}}, \ and\ \bibinfo {author} {\bibfnamefont
  {M.}~\bibnamefont {Saffman}},\ }\href@noop {} {\bibfield  {journal} {\bibinfo
   {journal} {Physical Review Letters}\ }\textbf {\bibinfo {volume} {96}}
  (\bibinfo {year} {2006})}\BibitemShut {NoStop}%
\bibitem [{\citenamefont {Johanning}\ \emph {et~al.}(2009)\citenamefont
  {Johanning}, \citenamefont {Braun}, \citenamefont {Timoney}, \citenamefont
  {Elman}, \citenamefont {Neuhauser},\ and\ \citenamefont
  {Wunderlich}}]{Johanning2009}%
  \BibitemOpen
  \bibfield  {author} {\bibinfo {author} {\bibfnamefont {M.}~\bibnamefont
  {Johanning}}, \bibinfo {author} {\bibfnamefont {A.}~\bibnamefont {Braun}},
  \bibinfo {author} {\bibfnamefont {N.}~\bibnamefont {Timoney}}, \bibinfo
  {author} {\bibfnamefont {V.}~\bibnamefont {Elman}}, \bibinfo {author}
  {\bibfnamefont {W.}~\bibnamefont {Neuhauser}}, \ and\ \bibinfo {author}
  {\bibfnamefont {C.}~\bibnamefont {Wunderlich}},\ }\href@noop {} {\bibfield
  {journal} {\bibinfo  {journal} {Physical Review Letters}\ }\textbf {\bibinfo
  {volume} {102}},\ \bibinfo {pages} {073004} (\bibinfo {year}
  {2009})}\BibitemShut {NoStop}%
\bibitem [{\citenamefont {Seidelin}\ \emph {et~al.}(2006)\citenamefont
  {Seidelin}, \citenamefont {Chiaverini}, \citenamefont {Reichle},
  \citenamefont {Bollinger}, \citenamefont {Leibfried}, \citenamefont
  {Britton}, \citenamefont {Wesenberg}, \citenamefont {Blakestad},
  \citenamefont {Epstein}, \citenamefont {Hume}, \citenamefont {Itano},
  \citenamefont {Jost}, \citenamefont {Langer}, \citenamefont {Ozeri},
  \citenamefont {Shiga},\ and\ \citenamefont {Wineland}}]{Seidelin2006}%
  \BibitemOpen
  \bibfield  {author} {\bibinfo {author} {\bibfnamefont {S.}~\bibnamefont
  {Seidelin}}, \bibinfo {author} {\bibfnamefont {J.}~\bibnamefont
  {Chiaverini}}, \bibinfo {author} {\bibfnamefont {R.}~\bibnamefont {Reichle}},
  \bibinfo {author} {\bibfnamefont {J.~J.}\ \bibnamefont {Bollinger}}, \bibinfo
  {author} {\bibfnamefont {D.}~\bibnamefont {Leibfried}}, \bibinfo {author}
  {\bibfnamefont {J.}~\bibnamefont {Britton}}, \bibinfo {author} {\bibfnamefont
  {J.~H.}\ \bibnamefont {Wesenberg}}, \bibinfo {author} {\bibfnamefont {R.~B.}\
  \bibnamefont {Blakestad}}, \bibinfo {author} {\bibfnamefont {R.~J.}\
  \bibnamefont {Epstein}}, \bibinfo {author} {\bibfnamefont {D.~B.}\
  \bibnamefont {Hume}}, \bibinfo {author} {\bibfnamefont {W.~M.}\ \bibnamefont
  {Itano}}, \bibinfo {author} {\bibfnamefont {J.~D.}\ \bibnamefont {Jost}},
  \bibinfo {author} {\bibfnamefont {C.}~\bibnamefont {Langer}}, \bibinfo
  {author} {\bibfnamefont {R.}~\bibnamefont {Ozeri}}, \bibinfo {author}
  {\bibfnamefont {N.}~\bibnamefont {Shiga}}, \ and\ \bibinfo {author}
  {\bibfnamefont {D.~J.}\ \bibnamefont {Wineland}},\ }\href@noop {} {\bibfield
  {journal} {\bibinfo  {journal} {Physical Review Letters}\ }\textbf {\bibinfo
  {volume} {96}},\ \bibinfo {pages} {4} (\bibinfo {year} {2006})}\BibitemShut
  {NoStop}%
\bibitem [{\citenamefont {Kim}\ \emph {et~al.}(2005)\citenamefont {Kim},
  \citenamefont {Pau}, \citenamefont {Ma}, \citenamefont {McLellan},
  \citenamefont {Gates}, \citenamefont {Kornblit}, \citenamefont {Slusher},
  \citenamefont {Jopson}, \citenamefont {Kang},\ and\ \citenamefont
  {Dinu}}]{Kim2005}%
  \BibitemOpen
  \bibfield  {author} {\bibinfo {author} {\bibfnamefont {J.}~\bibnamefont
  {Kim}}, \bibinfo {author} {\bibfnamefont {S.}~\bibnamefont {Pau}}, \bibinfo
  {author} {\bibfnamefont {Z.}~\bibnamefont {Ma}}, \bibinfo {author}
  {\bibfnamefont {H.~R.}\ \bibnamefont {McLellan}}, \bibinfo {author}
  {\bibfnamefont {J.~V.}\ \bibnamefont {Gates}}, \bibinfo {author}
  {\bibfnamefont {A.}~\bibnamefont {Kornblit}}, \bibinfo {author}
  {\bibfnamefont {R.~E.}\ \bibnamefont {Slusher}}, \bibinfo {author}
  {\bibfnamefont {R.~M.}\ \bibnamefont {Jopson}}, \bibinfo {author}
  {\bibfnamefont {I.}~\bibnamefont {Kang}}, \ and\ \bibinfo {author}
  {\bibfnamefont {M.}~\bibnamefont {Dinu}},\ }\href@noop {} {\bibfield
  {journal} {\bibinfo  {journal} {Quantum Information \& Computation}\ }\textbf
  {\bibinfo {volume} {5}},\ \bibinfo {pages} {515} (\bibinfo {year}
  {2005})}\BibitemShut {NoStop}%
\bibitem [{\citenamefont {Nagerl}\ \emph {et~al.}(1998)\citenamefont {Nagerl},
  \citenamefont {Bechter}, \citenamefont {Eschner}, \citenamefont
  {Schmidt-Kaler},\ and\ \citenamefont {Blatt}}]{Nagerl1998}%
  \BibitemOpen
  \bibfield  {author} {\bibinfo {author} {\bibfnamefont {H.~C.}\ \bibnamefont
  {Nagerl}}, \bibinfo {author} {\bibfnamefont {W.}~\bibnamefont {Bechter}},
  \bibinfo {author} {\bibfnamefont {J.}~\bibnamefont {Eschner}}, \bibinfo
  {author} {\bibfnamefont {F.}~\bibnamefont {Schmidt-Kaler}}, \ and\ \bibinfo
  {author} {\bibfnamefont {R.}~\bibnamefont {Blatt}},\ }\href@noop {}
  {\bibfield  {journal} {\bibinfo  {journal} {Applied Physics B-Lasers and
  Optics}\ }\textbf {\bibinfo {volume} {66}},\ \bibinfo {pages} {603} (\bibinfo
  {year} {1998})}\BibitemShut {NoStop}%
\bibitem [{\citenamefont {Knoernschild}\ \emph {et~al.}(2010)\citenamefont
  {Knoernschild}, \citenamefont {Zhang}, \citenamefont {Isenhower},
  \citenamefont {Gill}, \citenamefont {Lu}, \citenamefont {Saffman},\ and\
  \citenamefont {Kim}}]{Knoernschild2010}%
  \BibitemOpen
  \bibfield  {author} {\bibinfo {author} {\bibfnamefont {C.}~\bibnamefont
  {Knoernschild}}, \bibinfo {author} {\bibfnamefont {X.~L.}\ \bibnamefont
  {Zhang}}, \bibinfo {author} {\bibfnamefont {L.}~\bibnamefont {Isenhower}},
  \bibinfo {author} {\bibfnamefont {A.~T.}\ \bibnamefont {Gill}}, \bibinfo
  {author} {\bibfnamefont {F.~P.}\ \bibnamefont {Lu}}, \bibinfo {author}
  {\bibfnamefont {M.}~\bibnamefont {Saffman}}, \ and\ \bibinfo {author}
  {\bibfnamefont {J.}~\bibnamefont {Kim}},\ }\href@noop {} {\bibfield
  {journal} {\bibinfo  {journal} {Applied Physics Letters}\ }\textbf {\bibinfo
  {volume} {97}},\ \bibinfo {eid} {134101} (\bibinfo {year}
  {2010})}\BibitemShut {NoStop}%
\bibitem [{\citenamefont {Olmschenk}\ \emph {et~al.}(2007)\citenamefont
  {Olmschenk}, \citenamefont {Younge}, \citenamefont {Moehring}, \citenamefont
  {Matsukevich}, \citenamefont {Maunz},\ and\ \citenamefont
  {Monroe}}]{Olmschenk2007}%
  \BibitemOpen
  \bibfield  {author} {\bibinfo {author} {\bibfnamefont {S.}~\bibnamefont
  {Olmschenk}}, \bibinfo {author} {\bibfnamefont {K.~C.}\ \bibnamefont
  {Younge}}, \bibinfo {author} {\bibfnamefont {D.~L.}\ \bibnamefont
  {Moehring}}, \bibinfo {author} {\bibfnamefont {D.~N.}\ \bibnamefont
  {Matsukevich}}, \bibinfo {author} {\bibfnamefont {P.}~\bibnamefont {Maunz}},
  \ and\ \bibinfo {author} {\bibfnamefont {C.}~\bibnamefont {Monroe}},\
  }\href@noop {} {\bibfield  {journal} {\bibinfo  {journal} {Physical Review
  A}\ }\textbf {\bibinfo {volume} {76}} (\bibinfo {year} {2007})}\BibitemShut
  {NoStop}%
\bibitem [{\citenamefont {Noek}\ \emph {et~al.}(2013)\citenamefont {Noek},
  \citenamefont {Vrijsen}, \citenamefont {Gaultney}, \citenamefont {Mount},
  \citenamefont {Kim}, \citenamefont {Maunz},\ and\ \citenamefont
  {Kim}}]{Noek2013}%
  \BibitemOpen
  \bibfield  {author} {\bibinfo {author} {\bibfnamefont {R.}~\bibnamefont
  {Noek}}, \bibinfo {author} {\bibfnamefont {G.}~\bibnamefont {Vrijsen}},
  \bibinfo {author} {\bibfnamefont {D.}~\bibnamefont {Gaultney}}, \bibinfo
  {author} {\bibfnamefont {E.}~\bibnamefont {Mount}}, \bibinfo {author}
  {\bibfnamefont {T.}~\bibnamefont {Kim}}, \bibinfo {author} {\bibfnamefont
  {P.}~\bibnamefont {Maunz}}, \ and\ \bibinfo {author} {\bibfnamefont
  {J.}~\bibnamefont {Kim}},\ }\href@noop {} {\bibfield  {journal} {\bibinfo
  {journal} {Optics Letters}\ }\textbf {\bibinfo {volume} {38}},\ \bibinfo
  {pages} {4735} (\bibinfo {year} {2013})}\BibitemShut {NoStop}%
\bibitem [{\citenamefont {Hayes}\ \emph {et~al.}(2010)\citenamefont {Hayes},
  \citenamefont {Matsukevich}, \citenamefont {Maunz}, \citenamefont {Hucul},
  \citenamefont {Quraishi}, \citenamefont {Olmschenk}, \citenamefont
  {Campbell}, \citenamefont {Mizrahi}, \citenamefont {Senko},\ and\
  \citenamefont {Monroe}}]{Hayes2010}%
  \BibitemOpen
  \bibfield  {author} {\bibinfo {author} {\bibfnamefont {D.}~\bibnamefont
  {Hayes}}, \bibinfo {author} {\bibfnamefont {D.~N.}\ \bibnamefont
  {Matsukevich}}, \bibinfo {author} {\bibfnamefont {P.}~\bibnamefont {Maunz}},
  \bibinfo {author} {\bibfnamefont {D.}~\bibnamefont {Hucul}}, \bibinfo
  {author} {\bibfnamefont {Q.}~\bibnamefont {Quraishi}}, \bibinfo {author}
  {\bibfnamefont {S.}~\bibnamefont {Olmschenk}}, \bibinfo {author}
  {\bibfnamefont {W.}~\bibnamefont {Campbell}}, \bibinfo {author}
  {\bibfnamefont {J.}~\bibnamefont {Mizrahi}}, \bibinfo {author} {\bibfnamefont
  {C.}~\bibnamefont {Senko}}, \ and\ \bibinfo {author} {\bibfnamefont
  {C.}~\bibnamefont {Monroe}},\ }\href@noop {} {\bibfield  {journal} {\bibinfo
  {journal} {Physical Review Letters}\ }\textbf {\bibinfo {volume} {104}}
  (\bibinfo {year} {2010})}\BibitemShut {NoStop}%
\bibitem [{\citenamefont {Kim}\ \emph {et~al.}(2007)\citenamefont {Kim},
  \citenamefont {Knoernschild}, \citenamefont {Liu},\ and\ \citenamefont
  {Kim}}]{Kim2007}%
  \BibitemOpen
  \bibfield  {author} {\bibinfo {author} {\bibfnamefont {C.}~\bibnamefont
  {Kim}}, \bibinfo {author} {\bibfnamefont {C.}~\bibnamefont {Knoernschild}},
  \bibinfo {author} {\bibfnamefont {B.}~\bibnamefont {Liu}}, \ and\ \bibinfo
  {author} {\bibfnamefont {J.}~\bibnamefont {Kim}},\ }\href@noop {} {\bibfield
  {journal} {\bibinfo  {journal} {Ieee Journal of Selected Topics in Quantum
  Electronics}\ }\textbf {\bibinfo {volume} {13}},\ \bibinfo {pages} {322}
  (\bibinfo {year} {2007})}\BibitemShut {NoStop}%
\bibitem [{\citenamefont {Smith}\ \emph {et~al.}(1998)\citenamefont {Smith},
  \citenamefont {Rodgers}, \citenamefont {Sniegowski}, \citenamefont {Miller},
  \citenamefont {Hetherington}, \citenamefont {McWhorter},\ and\ \citenamefont
  {Warren}}]{Smith1998}%
  \BibitemOpen
  \bibfield  {author} {\bibinfo {author} {\bibfnamefont {J.}~\bibnamefont
  {Smith}}, \bibinfo {author} {\bibfnamefont {M.}~\bibnamefont {Rodgers}},
  \bibinfo {author} {\bibfnamefont {J.}~\bibnamefont {Sniegowski}}, \bibinfo
  {author} {\bibfnamefont {S.}~\bibnamefont {Miller}}, \bibinfo {author}
  {\bibfnamefont {D.}~\bibnamefont {Hetherington}}, \bibinfo {author}
  {\bibfnamefont {P.}~\bibnamefont {McWhorter}}, \ and\ \bibinfo {author}
  {\bibfnamefont {M.}~\bibnamefont {Warren}},\ }\href@noop {} {\bibfield
  {journal} {\bibinfo  {journal} {SPIE}\ }\textbf {\bibinfo {volume} {3514}},\
  \bibinfo {pages} {42} (\bibinfo {year} {1998})}\BibitemShut {NoStop}%
\bibitem [{\citenamefont {Knoernschild}\ \emph {et~al.}(2008)\citenamefont
  {Knoernschild}, \citenamefont {Kim}, \citenamefont {Liu}, \citenamefont
  {Lu},\ and\ \citenamefont {Kim}}]{Knoernschild2008}%
  \BibitemOpen
  \bibfield  {author} {\bibinfo {author} {\bibfnamefont {C.}~\bibnamefont
  {Knoernschild}}, \bibinfo {author} {\bibfnamefont {C.}~\bibnamefont {Kim}},
  \bibinfo {author} {\bibfnamefont {B.}~\bibnamefont {Liu}}, \bibinfo {author}
  {\bibfnamefont {F.~P.}\ \bibnamefont {Lu}}, \ and\ \bibinfo {author}
  {\bibfnamefont {J.}~\bibnamefont {Kim}},\ }\href@noop {} {\bibfield
  {journal} {\bibinfo  {journal} {Optics Letters}\ }\textbf {\bibinfo {volume}
  {33}},\ \bibinfo {pages} {273} (\bibinfo {year} {2008})}\BibitemShut
  {NoStop}%
\bibitem [{\citenamefont {Knoernschild}\ \emph {et~al.}(2009)\citenamefont
  {Knoernschild}, \citenamefont {Kim}, \citenamefont {Lu},\ and\ \citenamefont
  {Kim}}]{Knoernschild2009}%
  \BibitemOpen
  \bibfield  {author} {\bibinfo {author} {\bibfnamefont {C.}~\bibnamefont
  {Knoernschild}}, \bibinfo {author} {\bibfnamefont {C.}~\bibnamefont {Kim}},
  \bibinfo {author} {\bibfnamefont {F.~P.}\ \bibnamefont {Lu}}, \ and\ \bibinfo
  {author} {\bibfnamefont {J.}~\bibnamefont {Kim}},\ }\href@noop {} {\bibfield
  {journal} {\bibinfo  {journal} {Optics Express}\ }\textbf {\bibinfo {volume}
  {17}},\ \bibinfo {pages} {7233} (\bibinfo {year} {2009})}\BibitemShut
  {NoStop}%
\bibitem [{\citenamefont {Altepeter}, \citenamefont {Jeffrey},\ and\
  \citenamefont {Kwiat}(2006)}]{Kwiat2006}%
  \BibitemOpen
  \bibfield  {author} {\bibinfo {author} {\bibfnamefont {J.}~\bibnamefont
  {Altepeter}}, \bibinfo {author} {\bibfnamefont {E.}~\bibnamefont {Jeffrey}},
  \ and\ \bibinfo {author} {\bibfnamefont {P.}~\bibnamefont {Kwiat}},\
  }\enquote {\bibinfo {title} {Photonic state tomography},}\ in\ \href@noop {}
  {\emph {\bibinfo {booktitle} {Advances In Atomic, Molecular, and Optical
  Physics}}},\ \bibinfo {series} {Advances}, Vol.~\bibinfo {volume} {52},\
  \bibinfo {editor} {edited by\ \bibinfo {editor} {\bibfnamefont
  {P.}~\bibnamefont {Berman}}\ and\ \bibinfo {editor} {\bibfnamefont
  {C.}~\bibnamefont {Lin}}}\ (\bibinfo  {publisher} {Academic Press},\ \bibinfo
  {address} {Bellingham},\ \bibinfo {year} {2006})\ pp.\ \bibinfo {pages}
  {105--159}\BibitemShut {NoStop}%
\bibitem [{\citenamefont {Knill}\ \emph {et~al.}(2008)\citenamefont {Knill},
  \citenamefont {Leibfried}, \citenamefont {Reichle}, \citenamefont {Britton},
  \citenamefont {Blakestad}, \citenamefont {Jost}, \citenamefont {Langer},
  \citenamefont {Ozeri}, \citenamefont {Seidelin},\ and\ \citenamefont
  {Wineland}}]{Knill2008}%
  \BibitemOpen
  \bibfield  {author} {\bibinfo {author} {\bibfnamefont {E.}~\bibnamefont
  {Knill}}, \bibinfo {author} {\bibfnamefont {D.}~\bibnamefont {Leibfried}},
  \bibinfo {author} {\bibfnamefont {R.}~\bibnamefont {Reichle}}, \bibinfo
  {author} {\bibfnamefont {J.}~\bibnamefont {Britton}}, \bibinfo {author}
  {\bibfnamefont {R.~B.}\ \bibnamefont {Blakestad}}, \bibinfo {author}
  {\bibfnamefont {J.~D.}\ \bibnamefont {Jost}}, \bibinfo {author}
  {\bibfnamefont {C.}~\bibnamefont {Langer}}, \bibinfo {author} {\bibfnamefont
  {R.}~\bibnamefont {Ozeri}}, \bibinfo {author} {\bibfnamefont
  {S.}~\bibnamefont {Seidelin}}, \ and\ \bibinfo {author} {\bibfnamefont
  {D.~J.}\ \bibnamefont {Wineland}},\ }\href@noop {} {\bibfield  {journal}
  {\bibinfo  {journal} {Physical Review A}\ }\textbf {\bibinfo {volume} {77}}
  (\bibinfo {year} {2008})}\BibitemShut {NoStop}%
\end{thebibliography}%

\end{document}